\documentclass[conference]{IEEEtran}

\usepackage[utf8]{inputenc}

\usepackage{csquotes}

\usepackage[numbers]{natbib}

\usepackage{graphicx}
\graphicspath{{images/}}

\usepackage{amsmath}

\PassOptionsToPackage{hyphens}{url}\usepackage{hyperref}

\usepackage{array}

\usepackage{multirow}

\usepackage{float}

\begin{document}

\title{
    RESTRuler: Towards Automatically Identifying Violations of RESTful Design Rules in Web APIs
}

\author{
    \IEEEauthorblockN{
        Justus Bogner\IEEEauthorrefmark{1}, Sebastian Kotstein\IEEEauthorrefmark{2}, Daniel Abajirov\IEEEauthorrefmark{3}, Timothy Ernst\IEEEauthorrefmark{3}, Manuel Merkel\IEEEauthorrefmark{3}
    }
    \IEEEauthorblockA{
        \IEEEauthorrefmark{1}Vrije Universiteit Amsterdam, Amsterdam, The Netherlands, j.bogner@vu.nl
    }
    \IEEEauthorblockA{
        \IEEEauthorrefmark{2}Reutlingen University, Reutlingen, Germany, sebastian.kotstein@reutlingen-university.de
    }
    \IEEEauthorblockA{
        \IEEEauthorrefmark{3}
        University of Stuttgart, Stuttgart, Germany\\
    }
}

\maketitle


\begin{abstract}
RESTful APIs based on HTTP are one of the most important ways to make data and functionality available to applications and software services.
However, the quality of the API design strongly impacts API understandability and usability, and many rules have been specified for this.
While we have evidence for the effectiveness of many design rules, it is still difficult for practitioners to identify rule violations in their design.

We therefore present RESTRuler, a Java-based open-source tool that uses static analysis to detect design rule violations in OpenAPI descriptions.
The current prototype supports 14 rules that go beyond simple syntactic checks and partly rely on natural language processing.
The modular architecture also makes it easy to implement new rules.
To evaluate RESTRuler, we conducted a benchmark with over 2,300 public OpenAPI descriptions and asked 7 API experts to construct 111 complicated rule violations.

For robustness, RESTRuler successfully analyzed 99\% of the used real-world OpenAPI definitions, with some failing due to excessive size.
For performance efficiency, the tool performed well for the majority of files and could analyze 84\% in less than 23 seconds with low CPU and RAM usage.
Lastly, for effectiveness, RESTRuler achieved a precision of 91\% (ranging from 60\% to 100\% per rule) and recall of 68\% (ranging from 46\% to 100\%).
Based on these variations between rule implementations, we identified several opportunities for improvements.

While RESTRuler is still a research prototype, the evaluation suggests that the tool is quite robust to errors, resource-efficient for most APIs, and shows good precision and decent recall.
Practitioners can use it to improve the quality of their API design.
\end{abstract}


\section{Introduction}
Web Application Programming Interfaces (Web APIs) are the building blocks for realizing complex Web applications and for sharing data and functionality with other applications and software services~\cite{Jacobson2011}.
With Representational State Transfer (REST), there exists an architectural style for realizing Web APIs that is very popular in industry~\cite{Schermann2016,Bogner2019}. 
Introduced in 2000, ~\citet{restapi} positioned REST as a framework of constraints that modern Web applications, e.g., a Web API, must fulfill to achieve quality attributes like scalability, efficiency, and reliability~\cite{article:PrincipleDesignOfTheModernWebArchitecture:2002}.
Although these constraints rely on well-established Web standards like the Hypertext Transfer Protocol (HTTP) and Uniform Resource Identifiers (URIs) and despite its popularity, REST has never become a standard~\cite{article:Rodriguez2016}.
Instead, REST and its constraints describe the recommended behavior of a Web application but provide no concrete instructions on how to implement this behavior.

As this leaves room for different interpretations and design decisions when implementing Web applications~\cite{article:Rodriguez2016,conference:Renzel2012}, several works translated Fielding's REST constraints into best practices and rules, e.g.,~\cite{palma2017semantic,Richardson2007,Masse2011}.
These should instruct developers on how to implement a RESTful API and guide them toward a high-quality design.
For several RESTful API design rules, empirical studies have shown that they are perceived as important by practitioners~\cite{Kotstein2021} and also effective for increasing the understandability and therefore also the usability of Web APIs~\cite{Bogner2023}.
However, it is still difficult for practitioners to adhere to many of these rules and to identify rule violations in their design.
Not every practitioner has the same knowledge about these rules and how to adhere to them.
Also, many Web APIs have a large number of resources, which makes manual reviews at scale unfeasible.

We therefore introduce RESTRuler, a Java-based command line interface (CLI) tool that uses static analysis to detect and report design rule violations in OpenAPI documentation.
For our prototype, we have selected 14 design rules proposed by Massé~\cite{Masse2011}, for which empirical evidence for their importance and effectiveness has been provided~\cite{Kotstein2021,Bogner2023}.
RESTRuler has a modular architecture that allows its efficient extension with new rules, for which the corresponding algorithms for detecting rule violations have to be implemented.
Some implemented rules rely on simple string comparison methods, but others require more sophisticated natural language processing (NLP) techniques or custom machine learning (ML) models.

To ensure its quality, we have evaluated RESTRuler on 2,331 OpenAPI definitions of real-world Web APIs and 14 constructed OpenAPI definitions with complicated rule violations designed by 7 API experts.
Based on this benchmark, we evaluated RESTRuler's robustness (percentage of successfully analyzed OpenAPIs), performance efficiency (analysis duration plus RAM and CPU usage), and effectiveness (precision and recall per rule).
In this paper, we present RESTRuler, our reusable evaluation design, and the evaluation results, which show the promising capabilities of the tool.

\section{Background and Related Work}\label{sec:background}
This section discusses existing literature presenting best practices and rules for a RESTful API design. 
Furthermore, we review related work that proposes approaches for the automatic detection of rule violations in Web APIs.

\subsection{Best Practices and RESTful API Design Rules}
Several works translated Fielding's REST constraints into best practices and rules for guiding developers toward high-quality RESTful design.
Best practices and rules have been proposed in scientific articles~\cite{Pautasso2014,conference:Petrillo2016,palma2014detection} and textbooks~\cite{Richardson2007,book:RestInPractice:2010,Masse2011}.
Additionally, multiple studies analyzed the REST compliance among real-world Web APIs by comparing them against proposed best practices and REST constraints~\cite{conference:Renzel2012,article:Rodriguez2016,Neumann2018}.
Most studies concluded that only a few APIs are truly RESTful.

The assumption that practitioners do not judge all best practices and rules as important and perceive them differently in terms of their impact on the quality of a Web API was the motivation for our Delphi study~\cite{Kotstein2021}:
we confronted practitioners from different companies with 82 RESTful API design rules compiled by Massé~\cite{Masse2011} and let them rate their importance and impact on software quality.
Practitioners rated only 45 of 82 rules with high or medium importance.

In a follow-up experiment~\cite{Bogner2023}, we tested the effectiveness of several RESTful API design rules, i.e., if adhering to these rules significantly impacts the understandability of a Web API.
In detail, we presented 12 Web API snippets, where each snippet existed in two versions, one adhering to a specific design rule and one violating the rule, to 105 participants. 
The participants had to answer comprehension questions about each snippet and rated the perceived difficulty of understanding the snippet.
For the experiment, we selected nine design rules proposed by Massé~\cite{Masse2011} that were perceived as very important in~\cite{Kotstein2021}, plus three instances of another rule proposed by Richardson and Ruby~\cite{Richardson2007}.
We identified a significant negative impact on understandability for 11 of 12 rule violations.
Moreover, the study participants perceived 9 of 12 rule violations as significantly more difficult to understand.

\subsection{Studies Analyzing the Quality of Web APIs}\label{sec:RelatedWork}
Multiple studies analyzed the linguistic quality of Web APIs by using (semi-)automatic approaches for detecting patterns and their corresponding antipatterns.
\citet{palma2017semantic} presented the \textit{Semantic Analysis of RESTful APIs} (SARA) approach for the detection of linguistic patterns and antipatterns in Web APIs.
In detail, the authors defined 12 linguistic patterns and their corresponding antipatterns and proposed algorithms for their detection.
They implemented these algorithms as part of the \textit{Service-Oriented Framework for Antipatterns} (SOFA) introduced in~\cite{conference:moha2012} and tested them on 18 real-world Web APIs.
Most analyzed APIs used appropriate resource names and did not use verbs within URI paths.

SARA was also used by \citet{petrillo} to identify two linguistic patterns and their antipatterns in 16 Web APIs for cloud computing.
Another contribution of their work is a tool called \textit{CloudLex} for the extraction of lexicons of cloud computing APIs that allows the analysis of the terms used in the 16 Web APIs.
However, the lexicons of the APIs only shared a few terms, although they were all cloud computing APIs.
In another study, \citet{conference:Palma2021} investigated whether well-designed RESTful APIs are also well-designed from a linguistic perspective.
Using SOFA, the authors analyzed 8 Google APIs in terms of 9 design patterns and antipatterns, plus 12 linguistic patterns and antipatterns.
The study concluded that only negligible relationships existed between RESTful and linguistic design qualities.

In two follow-up studies, Palma et al. extended SOFA with further detection algorithms and applied the linguistic quality analysis on Web APIs of specific domains:
in~\cite{conference:Palma2022PuPaPr}, they assessed the linguistic design quality of private, partner, and public APIs.
For this, they tested more than 2,500 endpoints from 37 APIs against 9 linguistic patterns and antipatterns.
In~\cite{article:Palma2022iot}, they focused on Web APIs from the IoT domain.

\citet{BRABRA201965} presented an approach to support developers of cloud computing Web APIs in evaluating their management APIs in terms of compliance with REST and Open Cloud Computing Interface (OCCI) principles.
In addition to 24 OCCI patterns and antipatterns from their previous work~\cite{conference:Brabra2016}, they proposed semantic definitions of 21 REST patterns and antipatterns and specified detection rules in Semantic Web Rule Language (SWRL) syntax. 
For detected antipatterns, the approach recommends a set of corrections to comply with REST and OCCI principles.
The authors applied their approach to five real-world cloud computing Web APIs and evaluated its accuracy, usefulness, and extensibility.

\begin{figure*}[ht]
  \centering
  \includegraphics[width=\textwidth]{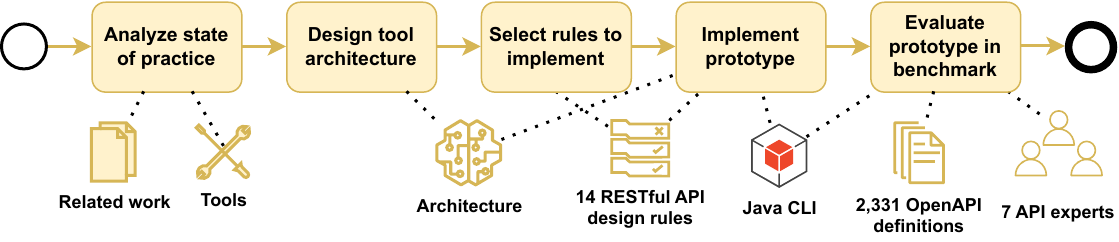}
  \vspace{-15pt}
  \caption{General Research Process}
  \label{fig:research-process}
\end{figure*}

While Palma et al.~\cite{palma2017semantic,conference:Palma2021,article:Palma2022iot,conference:Palma2022PuPaPr} and \citet{petrillo} focused on the detection of RESTful (anti-)patterns about URI design and proper HTTP verb use, the work of \citet{BRABRA201965} encompasses rules of various categories, like the correct use of HTTP headers, status codes, and hypermedia.
In our work, we focus mainly on rules addressing the design of URIs and the proper use of HTTP verbs but also cover rules related to metadata design and status codes.
While previous approaches have a small overlap with our implemented rules, there is so far no publicly available tool to conveniently check the empirically evaluated rules from \citet{Masse2011}.

\subsection{Practitioner Tools for Web API Quality Assurance}
The quality of Web APIs has not only been researched by academia, which we presented in the previous section, but is also an important topic in industry.
Several companies have shown that they take API quality seriously.
They have published written guidelines for implementing and documenting Web APIs and offered tools for validating APIs against these guidelines.
With Spectral\footnote{\url{https://github.com/stoplightio/spectral}}, Stoplight developed a popular linter that defines a basic set of rules that can be extended by individual rules and custom functions for the detection of violations in OpenAPI documents.
Companies like Adidas, Digital Ocean, IBM, Microsoft Azure, and Red Hat have translated their API guidelines into individual rule sets and implemented linters on top of Spectral.
IBM, for instance, has implemented 67 rules for their OpenAPI Validator.
Another tool named Zally\footnote{\url{https://github.com/zalando/zally}} has been implemented by Zalando. 
Zally supports the analysis of APIs based on the rules defined in Zalando's RESTful Guidelines.\footnote{\url{https://opensource.zalando.com/restful-api-guidelines/}}

During our analysis of existing API guidelines proposed by the aforementioned companies, we noticed that most proposed rules mainly target good practices for documenting APIs in OpenAPI and naming conventions rather than rules for a proper RESTful design.
Thus, for most rules, the detection relies on pure syntax analysis, e.g., the use of regular expressions to check naming conventions without any semantic verification.
For most of our envisioned rules, we would have to change the tool source code, thereby creating a fork.
This becomes especially evident in our evaluation section: we compare the performance of RESTRuler to Zally, the tool that supported the \enquote{most} of our selected rules: 4 of 14 rules.
We therefore decided to implement our own static analysis tool for OpenAPI descriptions to allow maximum flexibility in implementing the rules.

\section{Research Process}
The general process for our study is visualized in Fig.~\ref{fig:research-process}.
Initially, we analyzed both the state of the art, i.e., related work, and the state of practice, i.e., existing practitioner tools with a similar purpose (see Section~\ref{sec:background}).
In the next step, we designed the architecture for the tool, while taking inspiration from the conducted analysis.
Similarly to existing tools, we decided to start with a command line interface (CLI) that can be embedded into CI/CD pipelines or other applications.

We then made a selection of rules from the comprehensive catalog by \citet{Masse2011}.
This selection was guided by our extensive personal experience with these rules, the empirical evidence of our previous work (importance ranking~\cite{Kotstein2021} and understandability impact~\cite{Bogner2023}), and their implementation feasibility and estimated effort.
In detail, we first reduced Massé's catalog, consisting of 82 rules, to those that practitioners rated with high or medium importance~\cite{Kotstein2021}.
Afterward, we internally discussed which of these 45 remaining rules could be implemented and which could not.
For example, a rule could not be implemented if the detection of its violation requires information not provided in API descriptions.
Moreover, we estimated the effort and difficulty of implementing each rule and prioritized rules with empirical evidence for their effectiveness~\cite{Bogner2023}.
In the end, we selected 14 rules that we could implement in a realistic time: 10 of them were rated with high and 4 with medium importance in our Delphi study, and 6 of them showed a significant positive impact in our controlled experiment.

Afterward, we started the development of the CLI prototype based on rapid prototyping~\cite{devadiga2017tailoring}.
The goal was to quickly produce executable software with a minimum of functions, which was then iteratively refined based on feedback.
Three researchers were involved in the development, with the remaining two providing feedback and suggestions.
Additionally, a CI/CD pipeline was set up with GitHub Actions to build the CLI and run tests to avoid regressions during prototyping.
To implement individual rules, one researcher first created several rule violations that should be detected.
A second researcher then implemented the functionality for this, along with test cases and Markdown-based rule documentation.
Lastly, a pull request was created and reviewed by the third researcher before merging.
In the final step, we conducted an extensive benchmark-based evaluation of the implemented prototype that involved more than 2,300 public OpenAPI descriptions and 7 API experts to create custom rule violation examples.
The evaluation was guided by three research questions that correspond to the three properties we were interested in, namely \textit{robustness}, \textit{performance efficiency}, and \textit{effectiveness}.

\begin{enumerate}
    \item [\textbf{RQ1:}] How robust is the tool regarding the successful analysis of real-world APIs?
    \item [\textbf{RQ2:}] How fast and resource-efficient is the tool while analyzing real-world APIs?
    \item [\textbf{RQ3:}] How effective is the tool for correctly identifying rule violations?
\end{enumerate}

In the following, we first present details about RESTRuler and then discuss the design and results of its evaluation.

\section{Tool-Supported Approach (RESTRuler)}
RESTRuler\footnote{\url{https://github.com/restful-ma/rest-ruler}} is a static analysis tool to identify design rule violations in OpenAPI description files.
It is a CLI tool written in Java, which relies on \textit{Micronaut}\footnote{\url{https://micronaut.io}}, an open-source JVM-based framework for building applications with faster startup and lower memory requirements.
We also used \textit{picocli}\footnote{\url{https://picocli.info}} as a framework to simplify standard CLI functionality.
Below, we describe the standard workflow of the CLI, its architecture, and finally the implemented rules.

\subsection{CLI Workflow}
\begin{figure}[ht]
    \centering
    \includegraphics[width=\columnwidth]{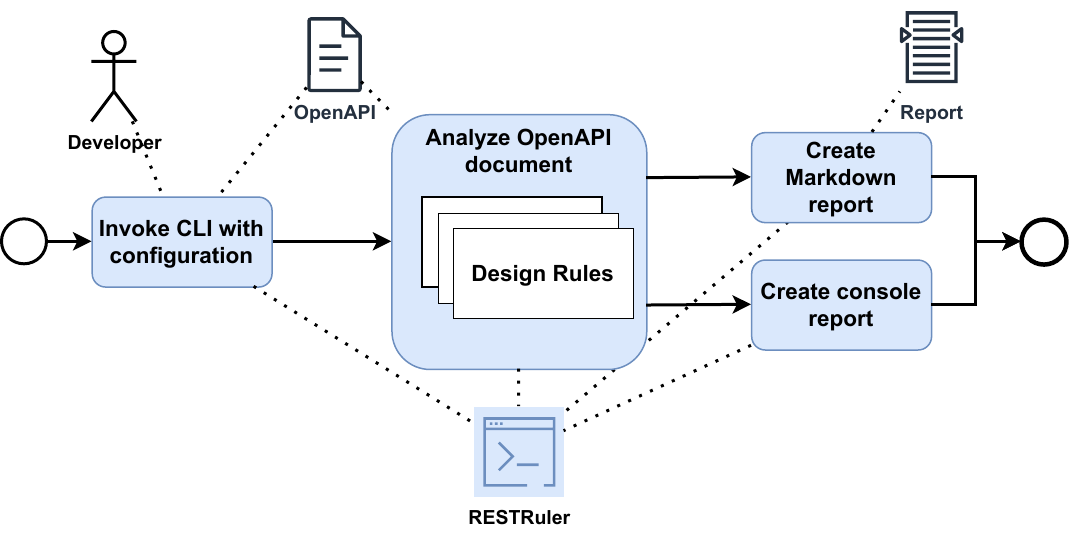}
    \vspace{-15pt}
    \caption{RESTRuler's CLI Workflow}
    \label{fig:cli-workflow}
\end{figure}

We visualize the general workflow of RESTRuler in Fig.~\ref{fig:cli-workflow}.
A developer interested in the design quality of their RESTful API invokes the RESTRuler CLI with the intended configuration.
The only mandatory parameter is the local path or remote URL to the OpenAPI description file, with a few additional parameters being optional, e.g., a flag to interactively select the rules that should be used for the analysis.
The tool then locates or downloads the OpenAPI document and parses it into an internal format.
RESTRuler now iterates through all active design rules and analyzes the API representation regarding violations of these rules, with found violations being saved for later.
After the analysis is finished, all identified rule violations are aggregated into a report that contains the concrete violations and their line numbers to quickly identify them in the file.
This report is always displayed as output for the console, but it can also be written into a Markdown file based on provided CLI parameters.
The Markdown report contains more details, e.g., rule category, severity, and a generic suggestion on how to fix the violation.
Due to this workflow, it is fairly easy to use RESTRuler manually but also to integrate it into automation scripts or CI/CD pipelines.

\subsection{Architecture}
\begin{figure}[ht]
    \centering
    \includegraphics[width=\columnwidth]{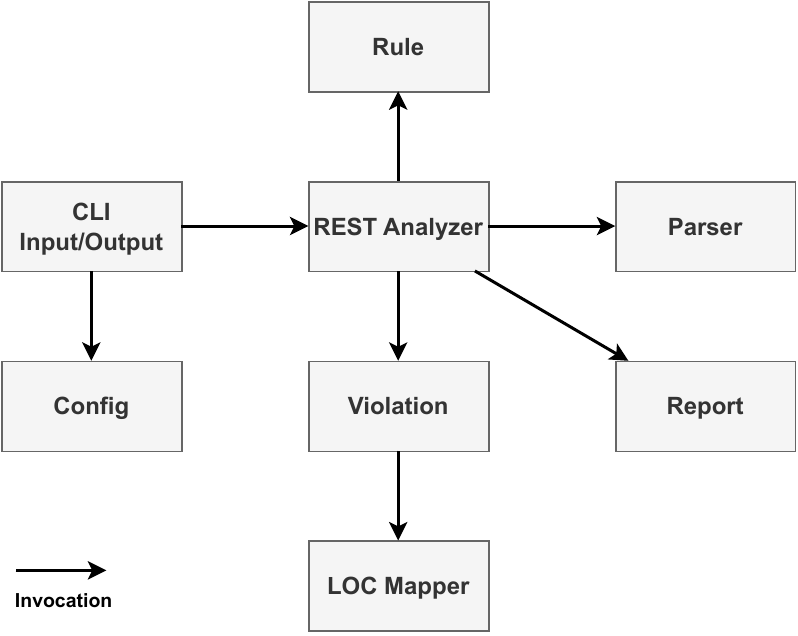}
    \vspace{-15pt}
    \caption{RESTRuler's Architecture of Modules and Their Dependencies}
    \label{fig:cli-architecture}
\end{figure}

We visualize the architecture containing the different modules that are necessary to achieve this workflow in Fig.~\ref{fig:cli-architecture}.
In the design, we placed special emphasis on modularity and extensibility, e.g., adding new rule implementations should be very efficient.
We briefly describe every module in the following listing.
\begin{enumerate}
    \item \textbf{\texttt{CLI Input/Output:}} This module is responsible for interfacing with the user. It controls the application flow, manages user input and CLI parameters, such as the path to the OpenAPI document, and also outputs messages and the contents of a report to the console.
    \item \textbf{\texttt{Config:}} This module is responsible for the configuration to be used for the analysis, such as the selection of active rules. This configuration is instantiated by the \texttt{CLI Input/Output} module and later provided to the \texttt{REST Analyzer}.
    \item \textbf{\texttt{Parser:}} This module reads the OpenAPI document in both YAML or JSON using the Swagger parser Java library.\footnote{\url{https://github.com/swagger-api/swagger-parser}} OpenAPI definitions in version 2.0 are automatically converted into comparable OpenAPI 3.0 definitions. The parser is fairly tolerant regarding schema violations, such as omitting content. The different parts of the parsed API can be conveniently accessed based on the Swagger data model.
    \item \textbf{\texttt{Rule:}} Each rule has to adhere to a standardized interface that needs to be implemented. Besides general attributes required for the reporting like a title or rule category, the backbone of the analysis is the \texttt{checkViolation()} method. An implementation needs to be provided that examines the parsed OpenAPI model and returns a list of all found violations.
    \item \textbf{\texttt{REST Analyzer:}} This module is the analysis orchestrator. The instantiated OpenAPI model is passed to each active rule for analysis. If a violation occurs, this information is forwarded to the \texttt{Violation} module. After a successful analysis, the collected violation objects are passed to the \texttt{Report} module.
    \item \textbf{\texttt{Violation:}} This module provides a unified representation of rule violations. It gets input from the \texttt{REST Analyzer} and uses the \texttt{LOC Mapper} to enrich the data. As output, it provides a violation object containing the respective API path, the line where the violation occurred, a description of the rule, and an improvement suggestion.
    \item \textbf{\texttt{LOC Mapper:}} To provide better feedback on the location of the violation, this module maps each path in the OpenAPI document to its line number.
    \item \textbf{\texttt{Report:}} This module is responsible for the final analysis output. It creates the simple summary report as output for the console, but it can also create the more detailed Markdown-based report using a Markdown generator library\footnote{\url{https://github.com/Steppschuh/Java-Markdown-Generator}}, and save it locally to a file.
\end{enumerate}

\begin{table*}[ht]
\caption{14 Implemented Rules from \citet{Masse2011}}
\label{tab:implemented-rules}
\centering
\begin{tabular}{lll}
    \textbf{Rule Description} & \textbf{Identifier} & \textbf{Category}\\
    \hline
    \hline
    A plural noun should be used for collection or store names & PluralNoun & URI Design \\
    A singular noun should be used for document names & SingularNoun & URI Design \\
    A verb or verb phrase should be used for controller names & VerbController & URI Design \\
    A trailing forward slash (/) should not be included in URIs & NoTrailingSlash & URI Design\\
    Forward slash separator (/) must be used to indicate a hierarchical relationship & ForwardSlash & URI Design\\
    File extensions should not be included in URIs & NoFileExtensions & URI Design\\
    CRUD function names should not be used in URIs & NoCRUDNames & URI Design \\
    Underscores (\_) should not be used in URI & NoUnderscores & URI Design \\
    Hyphens (-) should be used to improve the readability of URIs & Hyphens & URI Design \\
    Lowercase letters should be preferred in URI paths & Lowercase & URI Design \\
    Content-Type must be used & ContentType & Metadata Design \\
    \texttt{GET} and \texttt{POST} must not be used to tunnel other request methods & NoTunnel & Request Methods\\
    \texttt{GET} must be used to retrieve a representation of a resource & GETRetrieve & Request Methods\\
    \texttt{401 (Unauthorized)} must be used when there is a problem with the client's credentials & RC401 & HTTP Status Codes \\
    \hline
    \hline
\end{tabular}
\end{table*}

\subsection{Implemented Rules}
In total, we implemented 14 rules from \citet{Masse2011}.
The main focus was on rules from the category \textit{URI Design} (10 rules), but we also implemented one rule from \textit{Metadata Design}, two rules from \textit{Request Methods}, and one rule from \textit{HTTP Status Codes}.
We list these rules with an abbreviation and their category in Table~\ref{tab:implemented-rules}.
For a more in-depth description of the rule and how we implemented the violation detection, please refer to \citet{Masse2011} and our repository documentation.

The implementation of some rules like \textit{NoTrailingSlash}, \textit{NoUnderscores}, or \textit{Lowercase} relies on fairly simple string comparison methods in the URI path.
Other rules analyze the URI path based on dictionary lists, e.g., of certain unwanted words, sometimes in combination with regular expressions.
For example, \textit{NoFileExtensions} compares the end of each path segment against more than 800 common file extensions.
\textit{NoCRUDNames} does something similar based on a list of verb (sub-)strings that could represent CRUD operations, but also double-checks with a dictionary of more than 800 compound nouns in which these substrings would be acceptable.
\textit{Hyphens} uses a dictionary of 126k English words\footnote{\url{https://github.com/keredson/wordninja}} to identify if a path segment contains more than one word without a hyphen as a separator.
Other rule implementations analyze the existence or content of certain OpenAPI concepts to detect violations.
\textit{ContentType} examines response and request body specifications to verify that the content type is specified directly or in the components, except for \texttt{204 (No Content)} responses.
\textit{GETRetrieve} checks if a specified \texttt{GET} request falsely contains a request body and verifies that a response with status \texttt{200 (OK)} or a default response exists.
\textit{RC401} analyzes if all paths linked to a security scheme have a defined \texttt{401 (Unauthorized)} response.

Several other rules required more sophisticated natural language processing (NLP) capabilities, which we implemented via the Apache OpenNLP library\footnote{\url{https://opennlp.apache.org}}.
For example, \textit{SingularNoun} and \textit{PluralNoun} rely on the tokenization (\texttt{TokenizerME} and \texttt{SentenceDetectorME}) and Parts-of-Speech tagging (\texttt{POSTaggerME}) of this library.
In detail, we check whether a plural noun is used for a collection name and a singular noun is used for a document name for the respective URI concept, but also that these URI concepts alternate in the correct way, e.g., \texttt{<collection>/<document>/<collection>}, but not \texttt{<collection>/<collection>/<document>}.
Similarly, \textit{VerbController} checks whether a verb or verb phrase is absent within the last path segment of a \texttt{GET} or \texttt{POST} controller, which indicates a violation of this rule.

Lastly, \textit{NoTunnel} was one of the most complex rules with a strong semantic component. 
Tunneling refers to the abuse of \texttt{GET} or \texttt{POST} for a request that does not match the semantics of \texttt{GET} or \texttt{POST}, e.g., if \texttt{GET} is used to delete a resource.
To detect \textit{NoTunnel} violations in endpoints using \texttt{GET} or \texttt{POST}, the implementation analyzes the description and summary fields of the endpoint and compares their semantics with those of its HTTP verb.
For this, we use a custom machine learning model that predicts the correct HTTP verb based on description fields in natural language (or whether a description is invalid).
We trained this model with Weka\footnote{\url{https://waikato.github.io/weka-site/index.html}} on 3,227 manually labeled samples that we curated from public APIs.
Three researchers were involved in this process, with the dataset undergoing extensive review.
We chose a \textit{Naive Bayes Multinomial classifier} due to the potential vocabulary size of the descriptions, which could vary considerably.
To reduce dimensionality, the text data was preprocessed via text normalization, removing stop-words, and stemming.
We then applied Weka's \texttt{StringToWordVector} filter to transform text into a vector space.
Model performance was evaluated using 10-fold cross-validation, i.e., the dataset was randomly split into 10 subsets, each with roughly 323 instances.
In each cycle, the classifier was trained on 9 subsets (2,904 instances) and tested on the remaining 323 instances.
Our model achieved an average accuracy of 89\%.

\section{Evaluation Design}
\label{sec:evaluatio-design}
In this section, we describe the study design of our conducted benchmark with real-world APIs to evaluate RESTRuler's robustness (RQ1), performance efficiency (RQ2), and effectiveness (RQ3).
For transparency and reproducibility, we make all evaluation artifacts available online.\footnote{\url{https://doi.org/10.5281/zenodo.10246326}}

\subsection{Metrics}
We collected several metrics during our benchmark.
To operationalize \textit{robustness} (RQ1), we tracked whenever RESTRuler was not able to successfully analyze a real-world API and aborted the analysis with an error message.
As a final metric, we calculated the percentage of successfully analyzed OpenAPI documents, and expected it to be very close to 100\%.
For \textit{performance efficiency} (RQ2), we used the time it took to analyze each OpenAPI file, as well as the amount of used RAM and the percentage of used CPU during the analysis.
The goal was to have reasonable durations and resource usage that is acceptable for standard desktop PCs and laptops.
Lastly, for \textit{effectiveness} (RQ3), we used the two standard evaluation metrics for classification tasks, namely precision, i.e., the percentage of rule violations reported by the tool that were actually correct, and recall, i.e., the percentage of existing rule violations in a document that were correctly reported.
A perfect tool would achieve a precision and recall of 100\%, i.e., no false-positives (reported violations that are not actual violations) and no false-negatives (existing violations that were missed).
However, this is usually impossible to achieve for reasonably complex classifications, and such tools have to somehow ensure that both metrics are sufficiently high.
Tools that produce too many false positives are tiresome to use in practice, and developers will learn to ignore the reported violations, but tools that miss too many critical violations are also of dubious value.
For our initial prototype, precision was slightly more important than recall because it would encourage acting on the reported violations.

\subsection{Study Objects}
To assemble a large set of real-world OpenAPI descriptions for our study, we used \texttt{apis.guru}\footnote{\url{https://apis.guru}}, a public repository of OpenAPI specifications.
We used a script to download all OpenAPI definitions as JSON files, which resulted in a list of 2,331 distinct OpenAPI definitions.
These definitions strongly varied in size and covered a wide range of topics, with official cloud APIs from companies like Amazon and Microsoft to museum and government APIs.
There were also nearly no multiple versions of the same API, which added to the variety.
The documents also use OpenAPI concepts like components or responses very differently, making them an ideal testing ground for our implementation of API design rules.
To illustrate the size differences in our sample, and since it also makes the performance analysis more meaningful, we classified the OpenAPI documents into five size categories based on the number of paths: \textit{very small}, \textit{small}, \textit{medium}, \textit{large}, and \textit{very large}.
This categorization can be seen in Table~\ref{tab:CategorisationOfFiles}.
The number of paths per category was chosen based on the performance analysis of a small sample of different files.
We compared the needed time per path and then formed the buckets as approximations.
In our sample, the majority of APIs are fairly small: 62\% belong to the \textit{very small} category, 22\% to \textit{small}, 10\% to \textit{medium}, and only 4\% to \textit{large} and 2\% to \textit{very large}.

\begin{table}[H]
\caption{Size Distribution of the 2,331 Collected OpenAPI Files}
\label{tab:CategorisationOfFiles}
\centering
\begin{tabular}{lrr}
    \textbf{Size Category}& \textbf{Numbers of Paths} &\textbf{Number of OpenAPI Files}\\
    \hline
    \hline
    very small & 0-10  & 1,448 \\
    small & 11-30  &  514  \\
    medium & 31-70  & 234  \\
    large & 71-150  &  95  \\
    very large & 151+  & 40 \\
    \hline
    \hline
\end{tabular}
\end{table}

While this large set of real-world APIs is suitable for most defined metrics, it is impractical to use it for recall, simply because it is very difficult to identify the number of violations that RESTRuler missed (false-negatives).
Per analyzed API and rule, we would have needed to check for all missed violations, which is not feasible.
Extracting verified violations into a combined new dataset was also no option because we already knew that RESTRuler could identify these.
Therefore, we additionally asked seven software professionals familiar with REST and APIs to define custom violations for each rule.
All of them had at least several years of experience with Web APIs, either in industry or through research and teaching.
To further test the limits of RESTRuler, we asked them to also include some edge cases, i.e., rule violations that they perceived as difficult to detect with tools.
Before accepting the violations, their soundness was discussed within the research team.
In some cases, we sent clarifying questions to the respective expert.
The combined set of 111 rule violations was used as a gold standard for the recall evaluation, with one OpenAPI definition being created for each rule that contained all respective violations from the experts.

\begin{table*}[ht]
\caption{Performance Results}
\label{tab:PerformanceResults}
\centering
\begin{tabular}{lrrrrr}
    \textbf{Type}& \textbf{Median Duration} &\textbf{Median File Size} & \textbf{Median \# of Paths} & \textbf{Median Memory Usage} & \textbf{Median CPU Usage}\\
    \hline
    \hline
    very small & 7.25 s  & 0.03 MB & 3 & 497 MB & 8.4\%\\
    small & 23.52 s  &  0.14 MB  &  17 & 957 MB & 8.4\%\\
    medium & 54.73 s  & 0.36 MB  & 43 & 1067 MB & 8.4\%\\
    large & 117.77 s  &  0.69 MB  & 89.5 & 1111 MB & 8.4\%\\
    very large & 300.96 s  & 1.60 MB  & 206.5 & 1176 MB & 8.4\%\\
    \hline
    \hline
\end{tabular}
\end{table*}

\subsection{Data Collection}
To calculate the described metrics, we wrote a Python script to automatically run RESTRuler with each of the 2,331 OpenAPI definitions as input.
A standard laptop with a Ryzen 5 3600 6x 3.60GHz CPU and 16 GB RAM was used for this analysis.
For robustness, a second script identified all OpenAPI specifications that failed to be processed and did not generate a valid output file.
In a second iteration, we re-ran only the failed specifications to log and check their error messages.
Regarding performance efficiency, we followed the guidelines by \citet{5387059} for Java performance analysis.
As a result, we used the Python library \texttt{psutil}\footnote{\url{https://pypi.org/project/psutil}} for logging the RAM and CPU usage.
This library retrieves information about running processes and system load, which allows isolating the Java process responsible for the analysis.
Additionally, the duration of each analysis was recorded using the Python \texttt{time} module.
The results were saved in a CSV file.

Regarding precision, we wrote another script that randomly assigned 30 different reports to three researchers (90 in total).
Per OpenAPI report, a maximum of 30 reported violations was randomly chosen (or all violations if there were fewer than 30).
Each researcher then manually checked their assigned violations and documented if they were correctly classified by the tool.
Based on the true-positives (TPs) and false-positives (FPs) for the combined 1,596 checked violations, we then calculated the precision per rule.
However, for \textit{NoTrailingSlash} and \textit{VerbController}, which did not occur so often, we did not sample enough violations to include them in the manual analysis.
For recall, we simply let RESTRuler analyze the OpenAPI documents with the 111 custom violations from the API experts and then documented which violations were missed.
Using the TPs and false negatives (FNs), we then calculated the recall per rule.
Additionally, to have some comparison with existing tools, we also let Zalando's tool Zally analyze the same gold standard and calculated the recall.
However, since Zally only supports 4 of the 14 implemented rules, we could only compare these.

\section{Evaluation Results \& Discussion}
\label{sec:evaluation-results}
In this section, we present the results of the evaluation and discuss their interpretations and implications.

\subsection{Robustness (RQ1)}
RESTRuler's analysis of the 2,331 mined OpenAPI definitions successfully created a report file for 2,300 of them.
For 31 files, the analysis failed, leading to an overall success rate of 98.7\%.
Out of these 31 OpenAPI definitions, a single one failed due to a faulty definition and the remaining ones due to RAM issues caused by an excessive number of paths.
All of these 30 files belonged to the \textit{very large} category.
For word-related rules, the tool loads entries for every word found in the analyzed file from a dictionary, which can lead to high RAM utilization for APIs with a large number of very long paths.
In our tests, these crashes happened around a certain RAM threshold, i.e., something between 5 and 7 GB RAM, and also mostly for rules based on extensive text analysis like \textit{ForwardSlash}, \textit{PluralNoun}, \textit{SingularNoun}, or \textit{VerbController}.
Making the memory usage of these rules more efficient may allow the analysis of larger OpenAPI files in the future.

\begin{figure}[ht]
    \centering
    \includegraphics[width=\columnwidth]{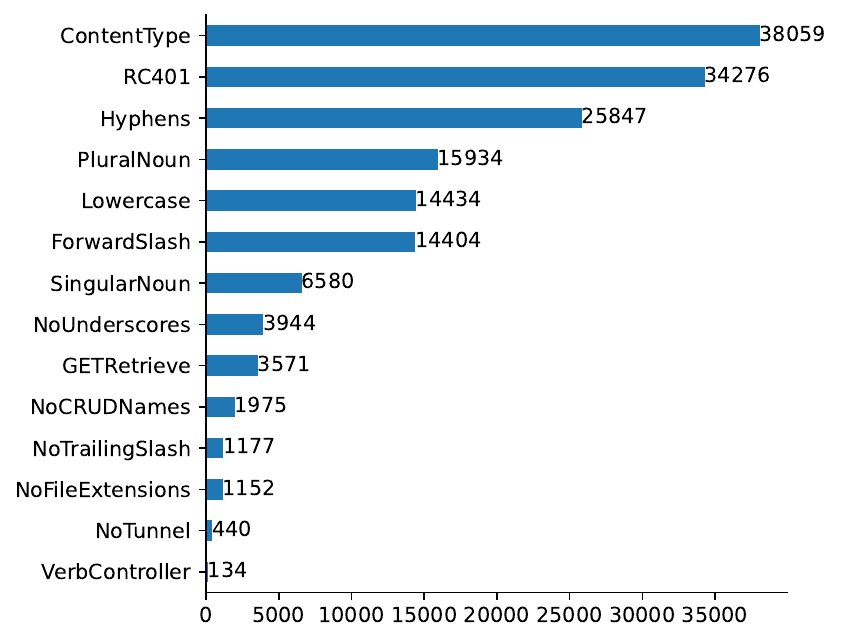}
    \vspace{-15pt}
    \caption{Distribution of the 169,061 Rule Violations in the Analyzed OpenAPIs}
    \label{fig:violations}
\end{figure}

Regardless, the overwhelming majority of the used real-world OpenAPI definitions were successfully analyzed.
In Figure \ref{fig:violations}, we can see the distribution of all 169,061 violations found in the 2,300 OpenAPI definitions, on average 73.5 violations per file.
Considering the percentage of fairly small APIs in our sample, this average seems quite large, but it was skewed by several outlier APIs with an excessive number of violations.
The median number of violations was 20 and the 75\% percentile was 69 violations.
Additionally, there are also several instances that led to multiple violations at once, e.g., \textit{NoUnderscores} or \textit{Lowercase} often appeared in combination with \textit{Hyphens} for the same path.
From a usability perspective, this is not ideal, and we plan to implement some sort of hierarchy within the rule set to make the violation reporting more lightweight.
Most notable is that \textit{ContentType} with 38,059 (22.5\%) and \textit{RC401} with 34,276 (20.3\%) were the most frequently violated rules, which together made up more than 40\% of the total.
Conversely, we saw only very few violations of \textit{VerbController} (0.08\%), \textit{NoTunnel} (0.26\%), \textit{NoFileExtensions} (0.68\%), and \textit{NoTrailingSlash} (0.70\%).

\subsection{Performance Efficiency (RQ2)}
We provide an overview of the performance evaluation results in Table~\ref{tab:PerformanceResults}. 
Regarding the analysis duration for all 14 rules, OpenAPIs in the category \textit{very small} with a median of 3 paths needed only around 7 seconds. 
As the number of paths increased, so did the analysis duration, e.g., files in the \textit{small} category (median of 17 paths) needed three times as much time (24 s).
For the \textit{medium} category, the required time increased to 55 seconds, which still seems acceptable for a manual analysis and especially CI/CD pipelines.
However, with the \textit{large} and \textit{very large} categories, the increased number of paths led to median execution times of around 2 min and 5 min respectively.
While this definitely does not allow a highly interactive manual analysis, it still seems reasonable in CI/CD pipelines.
Moreover, from a maintenance perspective, the number of paths in these categories is probably too large anyway and offers the potential for splitting.
These categories also only correspond to a small number of files in our sample (4.1\% and 1.7\%), with \textit{very small} to \textit{medium} making up the vast majority of real-world APIs.
Nonetheless, there are options to improve the performance of the dictionary- and NLP-based rules for large APIs.
We will look into implementing caching for analyzed path segments for these rules in the future.

While the execution duration scaled slightly better than linearly with the number of paths, we saw a different picture for the resource consumption.
For memory utilization (RAM), the largest difference was between the \textit{very small} category and the others, which were fairly close together.
Here, we had a median of 497 MB compared to 957, 1,067, 1,111, and 1,176 MB for the other categories.
CPU utilization was even more uniform and did not seem so relevant for the analysis.
Although there definitely were some peaks, the median was the same across all categories (8.4\%).
In summary, RESTRuler was able to analyze the majority of APIs (\textit{very small} to \textit{medium}) in a reasonable time, which scaled fairly linearly with the number of paths.
Memory usage increased considerably from \textit{very small} to \textit{small}, but then stayed fairly close to 1 GB in all categories (apart from some excessively large files that failed the analysis).
CPU usage does not play a strong role in the analysis and was comparable across all categories.
This means that standard laptop, desktop, or CI/CD server hardware is completely sufficient to run RESTRuler.

\subsection{Effectiveness (RQ3)}
We summarize the effectiveness results per rule separately for precision (Table~\ref{tab:precision-results}) and recall (Table~\ref{tab:recall-results}).
The recall results are also compared to Zally for the shared four rules.
For all rules combined, RESTRuler achieved a precision of 91.2\%, which ranged from 60\% to 100\% between individual rules, and a recall of 67.6\%, which ranged from 46.2\% to 100\%.
The worse recall can partly be explained by the experts' task to also think of difficult edge cases to really test the limits of automatic detection, but also by our general focus on keeping FPs limited.
Many of these edge cases might indeed be fairly rare in real-world environments.
Additionally, several missed expert violations were fairly easy to address by extending the rule implementation after the benchmark analysis, raising the overall recall to 75\% without much effort.

\begin{table}[ht]
\caption{Precision Results of RESTRuler (RQ3)}
\label{tab:precision-results}
\centering
\begin{tabular}{lrrr}
    \textbf{Rule}& \textbf{Reported Violations} &  \textbf{TPs} & \textbf{Precision}\\
    \hline
    \hline
    NoUnderscores & 20  & 20 & 1.000\\
    Lowercase &  140 & 140 & 1.000\\
    RC401 & 358  & 358 & 1.000\\
    NoCRUDNames & 21  & 20 & 0.952\\
    ForwardSlash & 202  & 192 & 0.951\\
    ContentType & 219  & 206 & 0.941\\
    PluralNoun & 176  & 160 & 0.909\\
    Hyphens & 285  & 234 & 0.821\\
    SingularNoun & 40  & 28 & 0.700\\
    NoTunnel & 86  & 58 & 0.674\\
    GETRetrieve & 9  & 6 & 0.667\\
    NoFileExtensions & 15  & 9 & 0.600\\
    NoTrailingSlash & 0  & -- & --\\
    VerbController & 0  & -- & --\\
    \hline
    \textbf{Total} & 1,596  & 1,456 & 0.912\\
    \hline
    \hline
\end{tabular}
\end{table}

\begin{table}[ht]
\caption{Recall Results of RESTRuler and Zally (RQ3)}
\label{tab:recall-results}
\centering
\begin{tabular}{lr|rr|rr|}
     & & \multicolumn{2}{c|}{RESTRuler} & \multicolumn{2}{c|}{Zally}\\
    \textbf{Rule}& \textbf{Violations} &  \textbf{TPs} & \textbf{Recall} &  \textbf{TPs} & \textbf{Recall}\\
    \hline
    \hline
    NoTrailingSlash & 2  &  2 & 1.000 & 2 & 1.000\\
    GETRetrieve & 8  & 8 & 1.000 & -- & --\\
    NoUnderscores & 4  & 4 & 1.000& -- & --\\
    Lowercase &  6 & 6 & 1.000& 6 & 1.000\\
    Hyphens & 9  & 8 & 0.889 & 2 & 0.222\\
    RC401 & 6  & 5 & 0.833& -- & --\\
    SingularNoun & 9  & 6 & 0.667& -- & --\\
    ContentType & 6  & 4 & 0.667 & -- & --\\
    NoFileExtensions & 8  & 5 & 0.625 & -- & --\\
    NoCRUDNames & 13  & 8 & 0.615& -- & --\\
    PluralNoun & 14  & 7 & 0.500& 14 & 1.000\\
    VerbController & 4  & 2  & 0.500& -- & --\\
    ForwardSlash & 9  & 4 & 0.444 & -- & --\\
    NoTunnel & 13  & 6 & 0.462 & -- & --\\
    \hline
    \textbf{Total} & 111  & 75 & 0.676& 24 & 0.774\\
    \hline
    \hline
\end{tabular}
\end{table}

Some rules with easily identifiable violations achieved a perfect score in precision and recall, e.g., \textit{NoUnderscores} or \textit{Lowercase}.
Due to the small sample size, \textit{NoTrailingSlash} was not included in the precision benchmark, but it is also easy to detect and achieved a recall of 100\%, albeit on only two violations.
Unsurprisingly, Zally also achieved a recall of 100\% for \textit{NoTrailingSlash} and \textit{Lowercase}.
Below, we discuss the remaining, more complex rules in more detail.

Another strongly performing rule was \textit{RC401}, which achieved an impressive 100\% precision on 358 violations, but missed 1 of 6 violations for recall, leading to only 83.3\%.
The miss was an edge case where the \texttt{401} response code was present, but incorrectly used \texttt{Forbidden} as the description instead of \texttt{Unauthorized}.
Similarly, \textit{GETRetrieve} achieved a recall of 100\% for 8 violations.
However, for precision, only 6 of the 9 reported violations were correct (66.7\%).
The three FPs heavily used nested references in the HTTP response, which RESTRuler did not identify as retrieval.
While these references could hinder understanding, they still resolve to the representation of a resource and are therefore no violations.

Some other rules achieved very good precision, but performed worse for recall.
For example, \textit{NoCRUDNames} managed to get a precision of 95.2\% over 21 violations, reporting only a single one incorrectly.
This FP was due to the word \enquote{updater} in a path segment, which RESTRuler falsely interpreted as a CRUD operation name.
The word was later added to an already existing list of allowed words containing CRUD operation names, which fixed the FP and increased precision to 100\%.
However, for recall, \textit{NoCRUDNames} only found 8 of 13 seeded violations (61.5\%).
The five FNs were due to synonyms of CRUD operation names provided by the experts that were not part of the rule's dictionary, e.g., \enquote{purge}, \enquote{fetch}, \enquote{retrieve}, or \enquote{add}.
After adding these synonyms to the rule implementation, recall was also increased to 100\%.

With 95.1\% over 202 reported violations, the rule \textit{ForwardSlash} also achieved very good precision.
The 10 FPs were due to paths containing parts that appeared to be separated by different symbols, e.g., a period as in \texttt{.../my-image.jpg}.
However, this is of course not a real path separation, but instead a violation of \textit{NoFileExtensions}.
Other cases were due to unusual special characters perceived as separators.
The achieved recall with 44.4\% over 9 violations could unfortunately not live up to the good precision.
All five FNs were concerned with the hierarchy part of the rule, which our current implementation does not support due to the implementation difficulty of this semantic relationship.
We plan to approach this in the future via a dictionary of word hierarchies or custom ML models, possibly taking inspiration from Palma et al.~\cite{palma2015restful,palma2017semantic}, who performed similar analyses.

While the already mentioned \textit{NoFileExtensions} rule was only present with 15 reported violations, it achieved the lowest precision of all rules, namely 60\%.
All six FPs were based on a structure that included \texttt{.../Microsoft.Sql/servers/...} in the path.
While this is definitely suboptimal, it is not a \textit{NoFileExtensions} violation, but a \textit{Hyphens} or \textit{ForwardSlash} one instead, depending on the designer's intent.
With 62.5\%, the recall was slightly better than other rules, but the tool missed 3 of the 8 expert violations.
One of these was due to an extension not included in the rule's dictionary (\texttt{.heic}) and the other two were not recognized because the extension was not separated by a dot, e.g., \texttt{.../html}.
Both issues were fixed after the benchmark, which increased recall to 100\%.

The \textit{Hyphens} implementation reported 285 violations and reached a precision of 82.1\%.
We identified a total of 51 FPs during our manual checks.
The recall with 88.9\% was even better, with only a single missed violation.
All FPs and FNs were due to words not present in the rule's dictionary.
Our selected vocabulary was a compromise between sufficient accuracy and performance efficiency.
During implementation, we also experimented with a much larger dictionary containing 300k more words.
While this improved accuracy, the analysis duration also increased substantially. 
One option for future refinements may be to give users the choice between the two dictionaries to allow a conscious tradeoff between accuracy and performance.
Surprisingly, Zally could only achieve a recall of 22.2\% due to many FNs.
One explanation could be that they might use an even smaller dictionary, or that Zalando might have simply heavily prioritized precision for this rule.

Another rule with very good precision was \textit{ContentType}, which achieved 94.1\% over 219 violations.
The 13 FPs were due to nested component references, which RESTRuler did not follow to the end.
For recall, the tool found 4 of the 6 seeded violations (66.7\%).
One of these was based on an incorrect content-type, and the rule only checks if the attribute is present, not if its value is semantically correct.
We plan to implement this part in future work.
The second FN was due to the rule not checking the content-type of path parameters, which we fixed afterward, thereby increasing recall to 83.3\%.

Intuitively, we would have expected similar results for the two related rules \textit{SingularNoun} and \textit{PluralNoun}, as they rely on the same NLP library.
However, \textit{SingularNoun} achieved a precision of 70.0\% over 40 violations, while \textit{PluralNoun} achieved 90.9\% over 176.
For recall, it was the other way around, with \textit{SingularNoun} achieving 66.7\% and PluralNoun only 50.0\%.
Implementing the logic for both rules is not straightforward and depends heavily on the used NLP library.
Most FNs and FPs were caused by instances with words that have the same singular and plural, e.g., \enquote{species}, that only have a singular, e.g., \enquote{information}, or that only have a plural, e.g., \enquote{jeans}.
Zally, on the other hand, achieved perfect recall for \textit{PluralNoun} (100\%).
However, it also produced many FPs during the recall benchmark, which are not visible in this evaluation.
It seems like Zalando prioritized recall over precision in their implementation of this rule.

As one of the more complex rules that relied on a custom ML model, \textit{NoTunnel} achieved a precision of 67.4\% over 86 and a recall of 46.2\% over 13 violations, the lowest recall of all rules.
Since the model uses the provided \enquote{summary} and \enquote{description} attributes, low-quality text in these fields can easily lead to FPs and FNs.
An ML-based detection approach will never reach 100\%, but there is definitely room for improvement.
For example, the seven FNs for recall were due to the tunneling not being clearly described in the description, but being explicitly modeled in the request body or via query parameters.
We plan to extend the rule implementation in the future to also take these attributes into account.

Finally, regarding \textit{VerbController}, not enough violations were reported for this rule (only 134 in total) to be included in the randomly drafted OpenAPIs, i.e., we did not check its precision.
One explanation for this might be that the concept of a controller in RESTful API design, i.e., a resource that cannot be represented as a typical CRUD operation and is therefore modeled with a verb phrase instead of nouns (e.g., \texttt{POST /restore-backup}), is not well known in industry.
For recall, the tool missed 2 of the 4 expert violations (50\%).
The FNs were caused by words that are both valid nouns and verbs, namely \enquote{present} and \enquote{permit}.
This ambiguity of the English language is difficult to overcome, and our experts cleverly exploited this fact with their custom violations.

\section{Threats to Validity}
Even though we carefully created our study design in several iterations and with feedback for refinements, there are still some threats to validity that we need to mention.

Regarding internal validity, several things could have potentially impacted the results.
We manually labeled 1,596 reported violations for the precision benchmark.
This work was split between three researchers, who carefully executed their tasks.
However, every violation was still checked by a single person, which leaves room for subjective bias.
Since most violations were very straightforward to check and the researchers stayed in close contact for questions during the labeling, we still believe that potential inconsistencies could not change the overall direction of the results.
To reduce individual subjective bias during the tool implementation, several researchers were involved, and each rule implementation was always tested by a second person and then reviewed by a third researcher.
Similarly, some violations created by the experts were incomplete, e.g., missing the \enquote{description} attribute, and also required translation into the OpenAPI format.
While we carefully discussed controversial or ambiguous violations in the extended research team and also asked the experts for confirmation in some cases, there is still a possibility that a few violations were not in the exact intended format of their creators.
Furthermore, the complete benchmark relied heavily on automation via scripts.
While we carefully checked the code and result plausibility, it is still possible that a few minor errors could have occurred, especially due to the study complexity, with the complete analysis of all OpenAPI definitions running for a total of two days.
We are confident that such potential errors did not substantially impact the aggregated results.

Regarding external validity, our precision benchmark was based on the OpenAPI repository \texttt{apis.guru}.
Many of the 2,331 included APIs are from large companies, e.g., public cloud providers.
The three companies AWS (5.7\%), Google (1.5\%), and Microsoft (0.3\%) alone account for a combined 7.5\% of all APIs, which definitely could limit generalization slightly. 
Additionally, we randomly selected a sample of 90 OpenAPI definitions, with a maximum of 30 random violations per file.
In total, we drafted 1,596 of the 169,061 violations (0.94\%).
Based on this, there is definitely a possibility that an unlucky random draw could have limited the variety of violation instances.
For several rules, the effectiveness results, especially for recall, were based on a small sample size and require additional research for confirmation.
We also have to be careful to generalize from the performance on the public APIs to private APIs inside company networks, as these might be different enough to lead to other results.
To counteract this, we also included manually created violations from API experts, who also took their industry experience into account.
However, the violations were still specifically created to test the rule implementations through complicated edge cases.
Therefore, it is possible that these might not be fully representative of an industrial environment.
Nonetheless, identifying these edge cases was important to increase the overall detection quality of RESTRuler.

Finally, RESTRuler is a static analysis tool that does not take runtime information into account.
While this could provide more detailed insights about API implementation and behavior, dynamic analysis is also much more complex because it requires a running API that often expects authentication.
Such tests can also impact the performance of the API, and need to be scheduled with great care.
It was therefore a conscious decision to focus on static analysis with RESTRuler.

\section{Conclusion}
To support practitioners with efficiently identifying RESTful design rule violations in OpenAPI definitions, we propose RESTRuler, a static analysis tool that currently implements 14 design rules by \citet{Masse2011}.
We evaluated RESTRuler's robustness, performance efficiency, and effectiveness with 2,331 public OpenAPI definitions and 111 custom violations created by API experts.
While RESTRuler remains a prototype, the collected data suggests that the tool is quite robust to errors and can efficiently analyze the majority of APIs.
However, excessively large APIs can lead to problems with RAM usage, as the dictionary-based rule implementations require a lot of memory.
For the effectiveness, RESTRuler reached a precision of 91\% (ranging from 60\% to 100\% per rule) and recall of 68\% (ranging from 46\% to 100\%).
The analysis, especially of the manually crafted expert violations, revealed many opportunities for improvements, several of which we already implemented to improve precision and recall even further.

For the future, we plan to continue refining and evolving RESTRuler based on the gained evaluation knowledge.
Improving individual rule performance, adding additional important rules, and increasing the tool's usability and configurability are some of the plans on the roadmap.
Additionally, we want to perform industry case studies with the tool to further evaluate and improve its effectiveness and perceived usefulness.
Nonetheless, RESTRuler can already provide decent value in the realm of API quality assurance.
To allow convenient reuse and to encourage adoption, we make the tool\footnote{\url{https://github.com/restful-ma/rest-ruler} and \url{https://doi.org/10.5281/zenodo.10246375}} and its evaluation artifacts\footnote{\url{https://doi.org/10.5281/zenodo.10246326}} publicly available.

\section*{Acknowledgment}
We kindly thank all seven API experts for using their valuable time to create custom rule violations for our benchmark!

\bibliographystyle{IEEEtranN}
\bibliography{references}

\end{document}